\documentclass[submission,copyright,creativecommons]{eptcs}
\providecommand{\event}{VPT 2016}

\usepackage{breakurl}             
\usepackage[utf8]{inputenc}
\usepackage{color}
\usepackage{caption}
\usepackage{enumitem}
\usepackage{listings}
\usepackage{amsmath}
\usepackage{amsfonts}
\usepackage{multicol}
\usepackage[all]{xy}
\usepackage{mathtools}

\definecolor{thisismygrey}{rgb}{0.5, 0.5, 0.5}

\newcommand{\cell}[2]{\langle #2 \rangle_\mathsf{#1}}
\newcommand{\codecell}[1]{\langle\; #1 \;\rangle_\mathsf{code}}
\newcommand{\bcell}[2]{\big\langle #2 \big\rangle_\mathsf{#1}}
\newcommand{\bigcell}[2]{\left\langle #2 \right\rangle_\mathsf{#1}}

\newcommand{\sms}[2][]{%
\small%
\begin{equation}\label{#1}%
#2
\end{equation}%
\normalsize}%

\newcommand{\smsnn}[1]{%
\small%
\begin{equation*}%
#1
\end{equation*}%
\normalsize}%

\title{Towards Trustworthy Refactoring in Erlang}
\author{
Dániel Horpácsi
\institute{Eötvös Loránd Univeristy\\Budapest, Hungary}
\email{daniel-h@elte.hu}
\and
Judit Kőszegi
\institute{Eötvös Loránd Univeristy\\Budapest, Hungary}
\email{koszegijudit@elte.hu}
\and
Simon Thompson
\institute{University of Kent\\Canterbury, U.K.}
\email{S.J.Thompson@kent.ac.uk}
}
\def\titlerunning{Towards Trustworthy Refactoring in Erlang}
\def\authorrunning{D. Horpácsi, J. Kőszegi \& S. Thompson}

\begin{document}
\maketitle

\begin{abstract}
Tool-assisted refactoring transformations must be trustworthy if programmers are to be confident in applying them on arbitrarily extensive and complex code in order to improve style or efficiency. We propose a simple, high-level but rigorous, notation for defining refactoring transformations in Erlang, and show that this notation provides an extensible, verifiable and executable specification language for refactoring. To demonstrate the applicability of our approach, we show how to define and verify a number of example refactorings in the system.
\end{abstract}




\section{Introduction}

If a user is to refactor their source code using a refactoring tool then they need to have confidence that the tool can be trusted. There are a variety of approaches to making refactoring tools more reliable and more trustworthy. Confidence  may be established by carrying out extensive testing of transformations and performing transparent changes, but complete guarantees can only be achieved by formal verification of refactoring correctness. Defining verifiable refactoring transformations is still a significant challenge.

Informally-specified refactorings are typically implemented as conditional transformations on abstract syntax trees; these trees contain details of every aspect of syntax, and so definitions using them are low-level and complicated, which in turn makes understanding and verifying the transformations difficult. If the abstraction level of the description of the refactoring is higher (representation-independent), then the definitions are more natural to read and write, as well as being more amenable to verification. 

In this paper we present a high-level formalism, which provides a simple but rigorous way to define conditional transformations. There is a large design space for transformation formalisations: we have set two design goals for the work here. First, we aim to narrow down the scope from generic program transformations to verifiable refactorings, and secondly, we aim to define refactorings for a particular programming language, namely Erlang. This second goal means that we can leverage users' knowledge of Erlang to make the descriptions more powerful and accurate, as well as letting us define executable and mechanically verifiable refactorings. 

The paper makes the following contributions:
\begin{itemize}
\item A simple, executable formalism for defining refactoring transformations for Erlang.
\item A design of a transformation formalism that reflects a particular programming language.
\item High-level refactoring schemes for verifiable extensive transformations.
\item A method for turning refactoring definitions into formally verifiable logic formulas.
\end{itemize}

The rest of the paper is structured as follows. In Section~\ref{sec:background}, we give a very brief overview on the program model and refactoring framework we work with. In Section~\ref{sec:defs}, we introduce how refactorings are defined in our formalism, while in Section~\ref{sec:verification} we show the methods we use to mechanically verify refactoring definitions. Section~\ref{sec:related} summarises the related work, and Section~\ref{sec:conclusion} discusses some further issues and concludes.


\section{Background}
\label{sec:background}

Our solution is designed to support the Erlang~\cite{CesariniThompson} programming language, while the program model we use is based on the concepts used in RefactorErl~\cite{referl}, a static analyser and refactoring tool for Erlang. This section gives a brief overview on the background and previous work we build our presentation upon.

\paragraph{Erlang.}

The refactoring language we present is in some aspects specific to its object language, Erlang. Erlang is a concurrent, impure, functional programming language. Programs written in Erlang are composed of files, which consist of a set of forms encapsulating series of expressions. Files define modules, and forms define program entities such as functions and records. Erlang is eagerly evaluated, and it is strongly but dynamically typed. Because of the dynamic nature of the language, it is rather challenging to provide static analysis and correct refactoring for its programs.

\paragraph{Program representation.}

Our solution supposes that the model -- the underlying program representation -- captures syntactic as well as semantic properties of code. In particular, the representation of a program is a (labelled, directed) semantic program graph~\cite{referl}, which is an extension of the abstract syntax tree with static semantic information. Each node has a unique identifier and thus our language handles nodes as references.

Semantic information is represented in terms of semantic nodes as well as links between syntactic and semantic units. For instance, a semantic node for a function  stores (in its label) the function's name, arity and whether it is pure or not, while it is connected to syntax tree nodes defining it or referring to it. The function is also linked to its containing module as well as to its call sites. When defining refactoring side-conditions, we build upon these semantic properties and connections.

\paragraph{Refactoring framework.}

The implementation of our language relies on the capabilities of the underlying refactoring system. Since RefactorErl makes sure that the appearance of the code is preserved, we only have to worry about behaviour preservation of transformations. Furthermore, our refactoring definitions omit the formalisation as well as the implementation of meta-theory and static analysis for Erlang, because these are provided by the framework~\cite{Toth:2011:SAC:2363407.2363416}.

The realisation exploits the various syntactic transformation and static analysis features present in RefactorErl. Indeed, the representation-dependent steps of the refactoring function execution are implemented by communicating with the underlying program model. For instance, the evaluation of semantic side-conditions is implemented as looking up specific labels and paths in the semantic program graph, while construction of new syntactic elements is carried out by instantiating an abstract syntactic pattern in the model (concrete syntactic elements and their formatting are handled by the framework).

\section{Refactoring Definitions}
\label{sec:defs}

In this section, we introduce the formalism in which we define verifiable refactoring transformations, i.e. proven-correct refactorings; we begin in Section \ref{rationale} with a rationale for the design of our definition formalism. In Section \ref{def-prime} we show how \emph{prime} refactorings are defined from scratch, including both \emph{local} and \emph{extensive} refactorings, as well as \emph{refactoring schemes}; we illustrate each of these features by a series of examples as we go. We conclude in Section \ref{def-composite} with a discussion of how \emph{composite} refactorings are described.

\subsection{Rationale}
\label{rationale}
%
The design goals of our 
language are the following:
\begin{itemize}
\item \emph{Intuitive:} there is no need for familiarity with term rewriting or static analysis.
\item \emph{Representation-independent:} only language-level concepts are used in the formalism (as opposed to program representation-level concepts such as abstract syntax nodes).
\item \emph{Verifiable:} definitions can be verified as being refactorings.
\item \emph{Executable:} definitions are not only specifications, but implementations as well.
\item \emph{Applicable:} enables defining a wide range of real-world refactorings.
\end{itemize}

\noindent
We made the following design decisions:
\begin{itemize}
\item \emph{Language-dependent:} restricting to a single target language, Erlang in this case, we are able to provide readability, ease of use and fidelity to the language.
\item \emph{Interpreted DSL:} we have implemented the formalism as an external domain specific language, so that the definitions are executed by an interpreter implemented in Erlang.
\end{itemize}

\paragraph{The smaller the better.}

Our approach builds upon the idea of defining refactorings in terms of a series of simpler, so-called \emph{micro-refactorings}~\cite{Opdyke:1992:ROF:169783}. Indeed, less complex definitions are easier to write and understand, and also they are more likely to be verifiable for semantics preservation. Moreover, sequencing already verified refactoring transformations into more complex ones obviously results in correct refactoring definitions.

\paragraph{Refactoring functions.}

We define refactorings as functions with parameters that may be Erlang values (such as numbers or strings) as well as references to program elements (represented by nodes of the semantic program graph). The return value is always a program element, a node reference of the same type as that of the refactoring target. As in Erlang, functions are identified by their name and arity; modules are not (yet) supported. Definitions are dynamically and loosely typed: implicit type conversions might happen between values and syntactic nodes of constants, and between nodes of semantic entities and their names. This makes it convenient to compose patterns and conditions, as values can be part of syntactic patterns, while program elements can be intuitively used as their associated value.

Transformation or refactoring? It is worth clarifying that we are giving a formalism for defining (conditional) program transformations. However, the formalism makes it possible to prove that the transformations are indeed refactorings, i.e. they will preserve the semantics of programs. It is possible to write non-refactoring transformations in the language, but they will not pass the verification phase. On the other hand, it can also happen that correct refactorings do not pass the verification phase as the proof system is only relatively complete; in this latter case, we do dynamic verification (see Section~\ref{sec:ver-app}).

\paragraph{The target program element.}

Refactorings, or program transformations in general, replace a program by a modified program. However, in practice, most of the code remains unchanged, only a few elements are modified, even though they may be relocated. Therefore, we do not define refactorings as transformations that rewrite a whole program, but as changes to particular syntactic elements. In particular, our refactoring functions always have an implicit parameter called \emph{THIS}, a reference to a node (a program element) in the model, resembling the implicit object parameter of method calls in OO languages. This node determines the focus and scope of the change made by the transformation.

If the refactoring is \emph{local} to a syntactic unit, the target should be set to the top of the subtree corresponding to the unit, which will be transformed according to the rule(s) specified in the definition. The refactoring definition is intended to only change the target node (and its corresponding subtree) without affecting other parts of the model. Target nodes may be semantic as well, depending on the refactoring definition. 

On the other hand, if the refactoring consists of simultaneous changes to various syntactic elements connected by semantic means, the target should be a semantic object (such as a variable or a function) represented by a \emph{semantic node} that groups together the syntactic nodes referring to the semantic unit. This implies that the changed part of the graph is determined by the tree rooted at the semantic node. For example, function renaming executed on a semantic function node transforms the definition clauses as well as the referring application expressions, all of which are syntactic units.

The concept of target nodes further simplifies refactoring definitions as well as their verification. They are not parametrised by values based on which the refactoring function determines its target node, since this functionality will be captured by the notion of node selectors.

\paragraph{Types of refactoring definitions.}

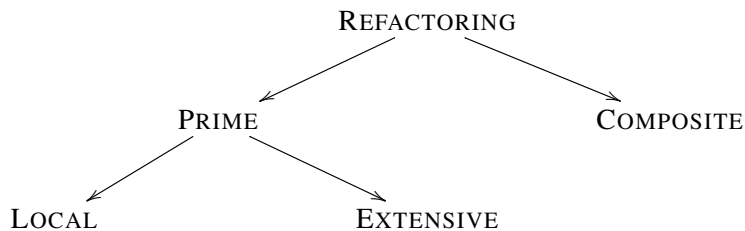
\begin{figure}[htb]
\begin{displaymath}
    \xymatrix{
        & & \textsc{Refactoring} \ar[dl] \ar[dr] &     \\
        &  \textsc{Prime} \ar[dl] \ar[dr]          &            & \textsc{Composite} \\
          \textsc{Local}           &  &        \textsc{Extensive} & }
\end{displaymath}
\caption{Types of refactoring definitions}
\label{fig:categ}
\end{figure}

Refactorings of different complexity are expressed at different abstraction levels, with different notation. Figure~\ref{fig:categ} shows the refactoring definition types we employ in our transformation formalisation.

Refactorings that cannot be expressed as a combination of other (smaller) refactorings are called \emph{prime}, while refactorings expressible as a series of other refactoring steps are called \emph{composite}. There might be different factorizations of composite refactorings. Prime refactorings are defined with conditional rewrite rules on syntactic program patterns, and combinations of these. Some refactorings can be expressed with a single rule, while others can only be defined as a combination of multiple rewrite rules. Refactorings of the former kind define shorter, module-local changes and are called \emph{local}, while the steps of the latter kind are called \emph{extensive}.

In the related work of formal refactoring definition, prime refactorings (or simple transformations in general) are mostly considered to be already defined on a lower level (e.g. with an API, outside the refactoring language) and are therefore called 'primitive' refactorings. In order to be able to verify complete refactorings, we specify even the simplest prime transformations inside our refactoring language. 

In order to simplify the definition and the verification of extensive transformations, we introduce \emph{refactoring schemes} that capture the general patterns underlying similar refactorings. These schemes can be instantiated with one or more conditional rewrite rules, and expand to refactoring transformations provided that the rewrite rules meet some constraints (verification issues are discussed in detail in Section~\ref{sec:ver-idiom}).

\subsection{Defining Prime Refactorings}
\label{def-prime}

\subsubsection{Local Refactorings}
\label{sec:localdef}

The simplest (typically local) refactorings can be specified by a single (conditional) rewrite rule on first-order syntactic terms. Note that the side-conditions of these transformations might be context-dependent, but their effect on the model is local to a program element (such as an expression, a function or a module).


\paragraph{Conditional rewrite rules} are the basis of refactoring definitions, providing a formalism for simple transformations of code fragments. They consist of a matching pattern, a replacement pattern and a conditions section. The patterns are specified with generalised program code, using concrete program syntax, which makes the patterns independent of the representation as well as the rules easy to read.

\begin{lstlisting}
REFACTORING <name> (<arguments>)
      <matching pattern>
    -----------------------
     <replacement pattern>
WHEN
  <conditions>
\end{lstlisting}

\noindent
Using this high-level notation for simple refactorings is not only intuitive, but it is easily verifiable as well, by building upon semantic equivalence of code patterns.

\paragraph{Example.}

Consider the following refactoring which simplifies list construction expressions by extracting the fragment defining the head of the list. If the head is given by a compound expression, it makes sense to simplify the expression by splitting it into two separate expressions by introducing a new variable. The condition guarantees that the introduced variable name (stored in metavariable \texttt{Var}) is a fresh, unbound name in the scope. Even though this is a context-sensitive condition, the transformation and its syntactic changes are local to the target expression.

\begin{lstlisting}[caption=Refactoring definition: extract\_listhead/0,label=lst:extractlisthead]
REFACTORING extract_listhead()
     [ HeadExpr | TailExpr ]
    -------------------------
     Var = HeadExpr,
     [ Var | TailExpr ]
WHEN
  fresh(Var)
\end{lstlisting}

\noindent
If the target node is not a top-level expression (that is, an element in the expression sequence of a clause), the result is automatically wrapped into a begin-end block --- we make use of this in the verification in Section~\ref{sec:ver-short}.

\paragraph{Patterns and metavariables.}

\textit{Patterns} are first-order terms expressed in concrete syntax, i.e. generalised syntactic terms involving metavariables that can match arbitrarily compound subterms. \emph{Metavariables} can be bound in two ways: they are matched against a syntactic subtree and get bound to the reference of the top node, or they are set by a condition attached to the rewriting rule. Metavariables are \emph{single-assignment}, they cannot be overwritten and therefore provide referential transparency, even across two or more rewrite rules combined. For the sake of simplicity, metavariables are denoted by Erlang variables; literal variables are matched by using a special semantic predicate.

Ordinary metavariables match exactly one syntactic subterm (subtree); however, there are special metavariables that can match zero, one or more consecutive, sibling subterms. These so-called \emph{list metavariables} are denoted by postfixing the variable name by two dots (\texttt{Args..}); not only can they be used in patterns, but can be bound in conditions as well (for example, when the result of a semantic function is a list rather than a single value). In addition, it is also possible to use multiple list metavariables in one pattern; this may result in multiple match results, but if the conditions do not narrow down the result set into exactly one solution, the matching fails.

In all prime refactoring definitions, the scope of a metavariable is the whole refactoring definition (even if it consists of multiple transformations). We will use this to allow combined rules to ``communicate'' via the metavariables used in the entire refactoring definition.

\paragraph{Example.}

The following refactoring definition demonstrates list metavariables. It matches simple (module-local) function applications and turns them into module-qualified (external) calls, making it explicit which module the called function belongs to. Since we match the arguments with a list metavariable, regardless of how many arguments the invoked function takes (zero or more), the expressions of actual parameters are simply reused in the new call.

\begin{lstlisting}
REFACTORING add_module_qualifier()
         Fun(Args..)
    -----------------
     Mod:Fun(Args..)
WHEN
  atom(Fun) AND Mod = module(THIS)
\end{lstlisting}



\paragraph{Semantic functions and predicates.}

Side-conditions of rewrite rules are usually specified by means of language-level concepts, e.g. ``$F$ is an exported function'', ``expression $A$ depends on expression $B$'' or ``expression $E$ is pure''. In our approach, such information is gathered via semantic functions and predicates, which are intended to cover all kinds of data available in the refactoring system, and is likely to be needed by refactoring definitions. Amongst others, there are semantic functions for querying properties of semantic entities such as modules, functions or variables, while predicates tell whether particular relationships exist between program units.

These functions are built-in and have a well-defined semantics, user-defined functions cannot be used in the conditions. The idea is somewhat similar to guards in Erlang: restrictions help give guarantees. When a rewrite rule is checked for being a refactoring, the rewrite patterns along with the conditions are transformed into a matching logic formula.


\paragraph{Rule conditions.} Rule conditions are first-order formulas built upon  semantic functions and predicates. Formulas are applications of semantic predicates, or structural equivalence checks on values of expressions; they are composed by negation, conjunction and disjunction. Expressions include constants, metavariables, as well as applications of semantic functions.

Formulas are evaluated left-to-right, call-by-value. This is important, because they may have side-effects: if the left-hand side of a matching condition (equality check) is an \emph{unbound} metavariable, the value of the right-hand side is bound to the metavariable. (Note that the semantics of this is very much similar to the match expression in Erlang, except that there is no pattern matching, only variables are allowed on the left.) Observe that metavariables bound this way can be used in the replacement pattern to contribute to the new subtree.

\paragraph{Example.}

The following example shows how matching conditions can be used to bind metavariables to results of semantic functions. The refactoring rewrites an Erlang list comprehension into an application of the \emph{map} higher-order function, whereas the generated list as well as the head function are extracted into auxiliary variables (\texttt{List} and \texttt{Fun}). The last expression in the result of this transformation might serve as target for a ``map to parallel map'' refactoring; thus, the composition of the two transformations would turn list comprehensions into parallel maps.

\begin{lstlisting}
REFACTORING listcomprehension_to_map()
    [ Head || GeneratorsFilters.. ]
   ---------------------------------------------
    List = [ {Vars..} || GeneratorsFilters.. ],
    Fun = fun({Vars.. }) -> Head end,
    lists:map(Fun, List)
WHEN
  Vars.. = intersect(bound_vars(GeneratorsFilters..), vars(Head)))
  AND fresh(List)
  AND fresh(Fun)
\end{lstlisting}

\noindent
Note that \texttt{Head} matches arbitrarily complex expressions, while \texttt{Vars..} captures all variables that are bound by the comprehension generators and are referred to in the comprehension head. The lists of variables returned by the semantic functions \emph{vars} and \emph{bound\_vars} are intersected according to set intersection; the ordering in the final result is undefined -- and irrelevant in this particular case.

\paragraph{Context-sensitivity.}

Although the pattern-based rewriting itself is context-insensitive, it is still possible to define seemingly complex, context-sensitive refactorings with single conditional rewrite rules. This is because the refactoring functions may receive context information in their parameters, and also, semantic predicates and functions are likely to return context-dependent data (for instance, in Listing~\ref{lst:extractlisthead}, the predicate \emph{fresh} states a context-dependent claim on the variable name). Note that 
we work with node references rather than terms, which is essential in querying context-dependent information on the various syntactic elements. Also, observe that not only can we make the side-conditions context-dependent, but via parameters and matching conditions, we can bind variables to context-dependent data and use them in the replacement.

\subsubsection{Extensive Refactorings}
\label{sec:extensivedef}

There are refactorings that cannot (practically) be expressed with a single rewrite rule. This is the case when the refactoring involves changes at multiple locations in the program, and the connection between these is purely semantic. Generally, such transformations are  only refactorings  if all the locations are changed at the same time, thus preserving consistency.

For example, if we rename a function at its definition, we need to change the name at all the reference sites as well, including directives, calls and other mentions. The connection between the elements to be changed is the semantic entity (the function in this case), the locations to be modified are determined by semantic relations such as ``defines'' and ``calls''. Also, this example demonstrates the typical scheme of extensive changes: there are some steps that make a twist in the semantics (changing a function name), which are then compensated by a series of additional changes (correcting the name at the call sites as well).

\paragraph{Combining rewrite rules.}

There are combinators in the language for composing rewrite rules into extensive transformations. Two well-known rewrite rule combinators have been adopted: \emph{sequencing (THEN)} and \emph{left-choice (OR)}. The semantics of these operators are basically the same as in Stratego~\cite{Bravenboer200852}: $A$~\textit{THEN}~$B$ executes $A$ first, and if it succeeds, executes $B$ too (if either $A$ or $B$ fails, all related changes are rolled back). In contrast, $A$~\textit{OR}~$B$ executes $A$, and proceeds to $B$ only if $A$ has failed for some reason.



\paragraph{Modifying rewrite rules.}

Although with combinators we can compose rewriting rules, without further modification, they will apply on the same part of the program, i.e. the target of the extensive refactoring function. We need additional operators to change the focus of the individual rewrite rules in the composition. Strategic rewriting solves this problem by using traversal operators that visit the children of the actual node. We introduce a more expressive notation: modifiers evaluate expressions that determine the nodes on which the rule applies.

The rewrite rules within extensive refactoring functions have their own target. By default, they inherit the target node of the refactoring, but with the following modifiers we can set different targets for the rewrite rule. The modifier \emph{ON} takes an expression, evaluates it (the result should be a node reference or a list of node references), and sets the target of the rule to the result. \emph{IN} is very similar, but the rule is applied not only on the result of the expression, but on all the nodes within its subtree.
Furthermore, there are modifiers for influencing the order in which the children of a node are visited.

\paragraph{Example.}

Consider renaming a function, which we have already used as an example for extensive change. The following definition shows a simplified version of the refactoring, which checks whether the new signature is free in the module and changes the name of the entity both in the defining clauses as well as at the simplest function calls. This transformation is a refactoring only if all the references are changed according to the modification in the definition.

\begin{lstlisting}
REFACTORING rename_function(NewName)
ON function_clauses(THIS)
       Name(Args..) -> Body..
   ---------------------------
    NewName(Args..) -> Body..
WHEN NOT function_exists(module(THIS), NewName, length(Args..))
THEN ON function_references(THIS)
       Name(Args..)
   -----------------
    NewName(Args..)
\end{lstlisting}

\noindent
In fact, the definition of the ``rename function'' refactoring would be much more complex than this, because there are a number of other ways to refer to a function entity in Erlang, such as module-qualified and apply calls, implicit fun expressions, export/import list entries, type and callback specifications. These all should be handled by such a ``function refactoring'' definition, because failing to modify the name at a reference will result in inconsistency: an incorrect reference.

In addition, there are many refactorings similar to ``rename function'', such as ``tuple function arguments'', ``reorder function arguments'' or ``add function argument'' -- what is common is the target, i.e. the function entity that is altered by the change. The function signature determines the name as well as the number and type of parameters the function takes. When we make a change in the signature, we have to carry out modifications at every site where the function is referred to. This leads us to a generic refactoring scheme, which covers all the function refactorings mentioned above. We look at that now.

\subsubsection{High-level Refactoring Schemes}
\label{sec:schemedef}

It is not easy to compose complete extensive refactoring steps, and in general, it is rather difficult to verify whether an extensive definition specifies a behaviour-preserving transformation, i.e. a refactoring. However, most extensive refactorings are in line with some change scheme: they alter a semantic entity such as a function, a variable, or a record, which has to be compensated by additional transformations.

In order to simplify the definition and verification of such refactorings, we introduce extensive change schemes and provide a basic set of them. These refactoring schemes can be instantiated with one or more rewrite rules (depending on the kind of the scheme). The instantiation results in a complex extensive refactoring transformation, which may rely on complex semantic properties (e.g. data flow or control flow) without the refactoring writer having to mention them explicitly. In some sense, the schemes can be considered as special strategies that check and process the rewrite rules passed to them.

There are two main benefits of using schemes: on one hand, extensive refactoring definitions become substantially simpler, and on the other, they become more easily verifiable. The schemes are built-in, they are proved to result in refactorings under some conditions; this is the contract of the scheme. If the instantiation is legal, the transformation is guaranteed to be a refactoring.

\paragraph{Function signature refactoring.}

All refactorings that change a function's signature have to change the definition as well as all the references, after checking whether there is no function with the new signature in the same module. The difference between these function refactorings is the way that they modify the name and the parametrisation of the function in question. Therefore, we provide a scheme for such refactorings, which captures all the general parts so that only the actual change in the signature has to be specified.

The function signature refactoring scheme makes it easy to define function-related refactorings: the parametrisation is a single rewrite rule defining the way the name as well as the arguments are changed. For example, the rename function refactoring becomes as simple as the following definition, and by using the very same scheme, another well-known Erlang refactoring can easily be defined, namely tupling the arguments of a function. Note that the side-conditions for these refactorings are defined by the scheme, ensuring that there exists no function with the new signature in the scope.

\begin{lstlisting}
FUNCTION SIGNATURE REFACTORING
  rename_function(NewName)
       Name(Args..)
   -----------------
    NewName(Args..)
\end{lstlisting}
\begin{lstlisting}
FUNCTION SIGNATURE REFACTORING
  tuple_function_arguments()
    Name( Args.. )
   ----------------
    Name({Args..})
\end{lstlisting}

\noindent
There are schemes for changing other semantic entities as well, such as \emph{modules} and \emph{records}. 

\paragraph{Forward dataflow refactoring.}

If we have a closer look at the function refactoring scheme, we might observe that the references to be changed with the definition are dependent on the definition: data and control (and therefore behavioural) dependencies are present between the referring expressions and the function definition. These dependencies induce the need for changing the program elements according to the same rule, at the same time.

This idea can be generalised, since such dependencies exist between various expressions, which means changing the one requires changing the others as well. Data dependencies are mainly caused by dataflow relations, so we provide a scheme for changing dataflow paths. If an expression constructing a value is changed, all the expressions into which the value flows (and therefore induces data and behavioural dependency) should be changed as well.

This skeleton is parametrised by a number of rules applied to either the construction site or a reference site of the data. That is, one of the definition rules is applied on the defining expression (the target of the refactoring), while the expressions referring to the data are transformed by one of the reference rules. In our current model, all elements on the dataflow path starting with the expression constructing the value are regarded as references. If the definition or any of the references cannot be transformed by a corresponding rule, the refactoring fails.

Note that there is an important side-condition for this scheme. Refactorings created with it will fail when any of the references to be compensated have any data sources (i.e. preceding dataflow nodes) other than the originally selected refactoring target. It is worth mentioning that if the target for this refactoring is the right-hand side of a match expression, and the matching pattern is a single, unbound variable, the previous conditions are apparently met.

\paragraph{Example.}

By instantiating the scheme, we can define a transformation eliminating the anonymous function wrapping a pure expression. The definition rule extracts the value, while the reference rules take care of the applications of the anonymous function. With a similar refactoring definition, we might inline the unnamed function by referring to the body of the function in the reference rules.

\begingroup
\begin{lstlisting}
FORWARD DATAFLOW REFACTORING fun2value()
DEFINITION
    fun() -> E end
   ---------------- WHEN pure(E)
             E
REFERENCE F
    F()
   -----
    F
REFERENCE G
    apply(G, [])
   --------------
          G
\end{lstlisting}
\vspace{-1.5em}
\captionof{lstlisting}{Forward dataflow refactoring example}
\label{lst:fwdata}
\endgroup

\medskip
\noindent
In many cases, the definition and reference rules are inverse in some sense: even in this case, this intuition helps understand the correspondence between the rules and their application. Let us see how this refactoring would change a simple code. Executing the ``fun2value'' refactoring on the fun expression checks if the value ``apple'' is side-effect free, and then it removes the unnecessary abstraction and application.

\vspace{1em}
\noindent
\begin{minipage}{.4\textwidth}
\begin{lstlisting}[mathescape,frame=single]
X = fun() -> apple end,
$\ldots$ ,
atom_to_list(X())
\end{lstlisting}
\end{minipage}
\quad{\Large$\xrightarrow{fun2value()}$}
\begin{minipage}{.4\textwidth}
\begin{lstlisting}[mathescape,frame=single]
X = apple,
$\ldots$ ,
atom_to_list(X)
\end{lstlisting}
\end{minipage}

\noindent
With this scheme, one might implement API-adaptation, or type-changing refactorings~\cite{Leather:2014:TRS:2543728.2543734} as well. Nevertheless, for such refactorings, data references should be gathered based on behavioural dependencies rather than just ordinary dataflow.


\paragraph{Backward dataflow refactoring.}

Changes in a dataflow path can be started from reference points as well, but with more restrictive preconditions: instances of the scheme can only be applied if the data sources of the selected expression do not flow anywhere but into the selected reference (this also means that the selected expression is the end of the dataflow path). If the selected expression is a control expression such that its data sources are its subexpressions, the condition is trivially met. Furthermore, if the refactoring copies nodes between the definition and the references, the names referred to by the copied units have to be common: in the following example, the unified tail may only refer to variables bound outside the \emph{case} expression.

\paragraph{Example.}

In this example, the constructed list is simplified into its head, while the tail is added to it after the next control flow node. Metavariables bound in the matching pattern of the definition rule but not used in the replacement thereof are treated as global in the transformation; this enables us to share \texttt{Xs} among the data sources and obligate data sources to have the same tail.

\begin{lstlisting}
BACKWARD DATAFLOW REFACTORING common_tail()
DEFINITION
    [X|Xs]
   --------
      X
REFERENCE Y
      Y
   --------
    [Y|Xs]
\end{lstlisting}

\noindent
Executing the above refactoring on the following case expression, the common tail is ``unified''.

\vspace{1em}
\noindent
\begin{minipage}{.4\textwidth}
\begin{lstlisting}[mathescape,frame=single]
f([H|T]) ->
    case H of
      1 -> [2|f(T)];
      3 -> [4|f(T)]
    end.
\end{lstlisting}
\end{minipage}
\quad{\Large$\xrightarrow{common\_tail()}$}
\begin{minipage}{.4\textwidth}
\begin{lstlisting}[mathescape,frame=single]
f([H|T]) ->
    [case H of
       1 -> 2;
       3 -> 4
     end       | f(T)].
\end{lstlisting}
\end{minipage}

\subsection{Defining Composite Refactorings}
\label{sec:compositedef}
\label{def-composite}

As we said in the beginning of the paper, the smaller the better, if it is about refactoring. Refactorings that are expressible as series of other refactorings should indeed be decomposed and specified with composite refactoring definitions.

The refactoring language has to help compose the already defined refactoring steps easily and safely. Note that even though composite definitions can be seen as extensive definitions that compose refactorings rather than just transformations (or, on the contrary, extensive definitions are compositions of transformations that are not all refactorings alone), we designed a separate formalism for composite definitions. The language enables easy and effective combination of refactoring definitions by allowing for defining node selectors and using the results thereof as target nodes for refactoring functions.

\paragraph{Do notation.}

The composite refactoring, basically, executes refactoring transformations defined by some control. It depends on the framework we use whether the steps are instrumented by branching and loop constructs, or are fired by non-deterministic choices and recursion. In our refactoring language, refactoring functions are applied to target nodes determined by metavariables and node selectors. By default, should any of the executed refactorings fail, the whole composition fails, and the changes made have to be rolled back.

By design, control is rather limited: unbounded recursion is not allowed in order to avoid non-termination; instead, the ON construct can be used to repetitively apply steps on a set of targets. Branching is also omitted, selectors are intended to implement conditional refactoring.

\paragraph{Selectors and executors.}

\emph{Selectors} are match-only functions which return node references without making any changes in the model. They can be used to collect potential targets for refactoring steps, as well as to gather context information passed to the refactoring function as a parameter. For example, the following selector matches functions taking at least one formal parameter, and returns the pattern expression belonging to the last parameter (the list metavariable \verb|Args..| matches all but the last).

\begin{lstlisting}
SELECTOR last_arg()
    Name(Args.., Last) -> Body.. .
RETURN Last
\end{lstlisting}

\noindent
\emph{Executors} provide a simple formalism for refactoring execution on nodes selected for transformation. When invoking a refactoring function, by default, its target is the target of the defining function. This can be overridden by targeting the function on specific nodes defined by selector expressions (including selector or semantic functions and metavariables). The modifier ``ON~$A$'' executes the refactoring function on the node(s) selected by $A$. In order to provide a more convenient formalism, a dot sign can shortcut the ON construct, i.e. \verb|A.refac()| is equivalent to \verb|refac() ON A|, resembling OO method invocations.

\paragraph{Example.}

Let us demonstrate the composition formalism by quoting a snippet from the ``generalise function'' refactoring definition. This is a fairly complex transformation that replaces a constant (or more generally, an expression) by a variable, which becomes a new parameter to the function. This refactoring produces a more general variant of the target function. We decomposed this transformation into multiple simpler refactorings.

The below definition creates a copy of the original function and rewrites it to refer to the newly generalised version. There are three main operations carried out: the target expression is wrapped into an anonymous function (as it can have side effects), an identical but generalised copy of the containing function is created, and finally, the copy is folded against the original definition as well as the generalisation is performed in the original definition by replacing the value with the new parameter.

\begin{lstlisting}[mathescape]
REFACTORING generalise_function()
DO
    OrigName = name(function(THIS))
    Orig     = function(THIS)

    THIS.wrap_into_fun()
    FunExp   = THIS.fun_part()

    Copy     = Orig.copy_function('tmp_name')
    Copy.add_parameter()
    Copy.rename_function(OrigName)

    LastArg  = definition(Copy).last_arg()
    Copy.fold_entire_function(Orig, copy(FunExp))
    FunExp.replace_val_by_var(copy(LastArg))
\end{lstlisting}


\noindent
We omit the definitions of the constituent refactorings, as our goal with this example is to demonstrate the composition mechanism. Nevertheless, we refer to refactoring steps such as wrapping an expression into an anonymous function (\verb|wrap_into_fun/0|), copying a function (\verb|copy_function/1|) and folding a function body against another function that has an identical body (\verb|fold_entire_function/2|). Interestingly, some of these helper refactorings are also decomposable into even smaller refactoring steps.

Note that since consecutive refactoring steps might depend on each others' result, we store the results of some refactoring functions into local variables: in this case, \texttt{Orig} captures the generalised function object, while \texttt{Copy} is a reference to the copy of the function, which is being further transformed.

\section{Verification}
\label{sec:verification}

By \emph{correctness} for a \emph{refactoring definition} we mean that the refactoring preserves behaviour when applied to any program of the object language. In this section we propose a verification technique that is suitable for verifying local refactorings as well as extensive refactorings expressed with schemes. Since the suggested proof system is not complete, we also address how to use a similar method for proving the correctness of an \emph{application} of the refactoring, i.e. that the original and the transformed code are equivalent.

A great advantage of our formalism is that a refactoring definition is an \emph{executable specification}, so we can reason about the transformation directly avoiding the usual gap between the specification and the implementation.
To be able to formally verify refactorings, besides the formal specification of the refactoring, we need to have 1) the formal semantics of the object language; 2) the formalization of the semantic properties used in the conditions of the transformations; 3) a logic into which the language semantics can be easily embedded as well as in which the behaviour-preservation property can be expressed; 4) a proof system for the logic.

The recently introduced \emph{reachability logic}~\cite{one-path-rl} (RL) can be a suitable all-in-one solution: it allows us to define the operational semantics of programming languages as well as to specify and to reason about program properties. The overall idea of the proposed approach is to define the semantics of Erlang in terms of RL formulas, expressing the correctness property of a refactoring as an equivalence problem. Then, by reducing partial equivalence to partial correctness according to Ciobaca~\cite{eq-corr}, we become able to use the proof system for RL to verify refactorings.

\subsection{From Reachability Logic to Equivalence Checking}
\label{sec:ver-rl}

As reachability logic is not a mature, well-known logic, its definition and the corresponding proof systems have been constantly evolving in various publications from recent years. In this section, a brief introduction is given to the related results~\cite{sym-ex, eq-corr} on which we build our verification technique.

\paragraph{Matching logic.} Reachability logic builds upon \emph{matching logic}, which is a specialized many-sorted first-order logic with a distinguished sort \texttt{Cfg}, called \emph{configuration}; additionally, it allows configuration terms with variables (called \emph{basic patterns}) as predicates. Let $T_{\texttt{Cfg}}(\text{Var})$ be the set of basic patterns, that is, terms of sort \texttt{Cfg} over the variables \texttt{Var}.

A basic pattern is satisfied by all the configurations that match it. Formally, the matching logic satisfaction relation $\models$ can be defined inductively as in first-order logic, extended with the following for basic patterns $\pi \in T_{\texttt{Cfg}}(\text{Var})$: $(\gamma,\rho) \models \pi\; \text{iff}\; \rho(\pi) = \gamma$, where $\gamma$ is a configuration and $\rho$ is a valuation.

Program states are represented as concrete configurations (ground configuration terms), while program state specifications are represented as \emph{patterns}, that is, first-order logic formulas with basic patterns. We use a special subset of patterns, called \textit{pure patterns}, for defining the operational semantics of the Erlang language as well as for expressing program pattern equivalence relations, because they can be ported directly to the $\mathbb{K}$ semantic framework we used for our semi-automatic method (see:~\ref{sec:auto}). A pattern $\varphi$ is pure~\cite{eq-corr} (or elementary~\cite{sym-ex}) if it is given in the form $\pi \wedge \varphi'$, where $\pi$ is a basic pattern and $\varphi'$ is a simple first-order logic formula without any basic pattern, called the condition of the pattern.
 
\paragraph{Configurations.} The configuration is usually a nested structure of cells containing semantic data. We will use the following simplified configuration for Erlang program states: $\cell{cfg}{\cell{code}{\ldots}\; \cell{env}{\ldots}\; \cell{defs}{\ldots}\;}$, where $\cell{cfg}{\ldots}$ is a top-level container cell, $\cell{code}{\ldots}$ contains the code to be executed, $\cell{env}{\ldots}$ stores variable assignments in a map, and $\cell{defs}{\ldots}$ contains function definitions. The following is a concrete configuration:

\vspace{-1em}
\smsnn{\cell{cfg}{\codecell{[\texttt{f(X)}\; |\; [2,3]]}\; \cell{env}{\texttt{X} \mapsto 1}\; \cell{defs}{\texttt{f}(\texttt{A}) \rightarrow \texttt{A} + 1}}}

\vspace{-1em}
\noindent
A pure pattern that is satisfied by the above concrete configuration would be:
\smsnn{\cell{cfg}{\codecell{[h\; |\; t]}\; \cell{env}{e}\; \cell{defs}{d}}\; \land\; \texttt{length}(t) > 0}

\vspace{-1em}
\noindent
Note that $\texttt{X}$ and $\texttt{A}$ are Erlang program variables represented as constants in matching logic, whereas $h, t, e$ and $d$ are mathematical variables.

\paragraph{Reachability logic.} While matching logic is a logic of static configurations, reachability logic is a logic of pairs of configurations representing dynamic behaviour: with RL one can express semantics of a programming language, as well as program properties. Given two matching logic formulas $\varphi$ and $\varphi'$, one can construct the reachability rule $\varphi \Rightarrow \varphi'$ stating that a configuration matching $\varphi$ will advance into a configuration matching $\varphi'$.

\paragraph{Matching logic semantics.} For the matching logic semantics of a language we need to define the semantic domain and a set of reachability rules capturing the operational semantics of the language. In practice, defining the semantic domain means to specify the abstract syntax of the programming language as well as the syntax of the operations in the needed mathematical domains, and give the model of configurations merged together with the mathematical domains. 

The object language of our refactorings is Erlang, but presenting the matching logic semantics for the entire language goes beyond the scope of this paper. Nevertheless, we show some example semantic rules in the following, and for the sake of simplicity, we use the above shown configuration sort. In our proof of concept, we have defined a deterministic, pure, single-module variant of Erlang. Due to the modular nature of matching logic semantics, we can easily extend this language by adding new cells and new reachability rules.
Even though Erlang does not have an official formal semantics definition, the user manual offers sufficient informal description about the meaning of language elements. Besides, we used some ideas from the doctoral thesis of Fredlund~\cite{fredlund-thesis}, which defines a small-step operational semantics for Erlang.

Let us show a formula defining the semantics of a begin-end block with an expression sequence beginning with a match expression (a similar definition is given in~\cite{fredlund-thesis}). We use ellipsis for the irrelevant and unchanged parts of the configurations.
\sms[eq:r1]{
\begin{aligned}
&\cell{cfg}{\codecell{\texttt{begin}\; \textit{pat} = \textit{exp},\; \textit{exps}\; \texttt{end} \; \ldots}\; &&\ldots} \land \texttt{length}(\textit{exps}) > 0\; \Rightarrow \\
&\cell{cfg}{\codecell{\texttt{case}\; \textit{exp}\; \texttt{of}\; \textit{pat}\; \texttt{->}\; \texttt{begin}\; \textit{exps}\; \texttt{end}\; \texttt{end}\; \ldots}\; &&\ldots}
\end{aligned}
}%

A more complex formula shows one of the four semantic rules for \emph{case} expressions. The predicate \verb|isMatching| checks whether the expression matches the pattern with respect to the variable environment, while \verb|getMatching| returns the new variable assignments resulting from the match. The function \verb|substVars| substitutes variables by their values in the given expression.
\sms[eq:r2]{
\begin{aligned}
&\cell{cfg}{\codecell{\texttt{case}\; \textit{exp}_1\; \texttt{of}\; \textit{pat}\; \texttt{->}\; \textit{exp}_2\; \texttt{end}\; \ldots}\; 
 &&\!\cell{env}{e}\; \ldots} \\
&\qquad\qquad \land \texttt{isMatching}(\textit{exp}_1, \textit{pat}, e) \land e_2 = \!\!\!\! && \texttt{getMatching}(\textit{exp}_1, \textit{pat}, e) \Rightarrow \\
&\cell{cfg}{\codecell{\texttt{substVars}(\textit{exp}_2, e_2)\; \ldots}\; 
 &&\!\cell{env}{e}\; \ldots}
\end{aligned}
}%

\paragraph{Program equivalence.}

In several cases we are able to derive the correctness of a refactoring to an equivalence problem. However, in the general-purpose proof systems for RL we can only reason about a property of a single program, but not about a relation of two programs. Ciobaca~\cite{eq-corr} shows that the problem of establishing partial equivalence can be reduced to the problem of showing partial correctness in a mechanically constructed aggregated language. (Partiality in this case means that we can only prove the equivalence of terminating programs, but most of our refactoring definitions remain in scope.) In our case, the configuration for the aggregated language is a pair of single program configurations:
{\small
\begin{equation*}
\bcell{eq}{\;\cell{cfg1}{\cell{code}{\ldots}\; \cell{env}{\ldots}\; \cell{defs}{\ldots}\;}\quad \cell{cfg2}{\cell{code}{\ldots}\; \cell{env}{\ldots}\; \cell{defs}{\ldots}\;}\;}
\end{equation*}
}%

For each semantic rule, we have to generate two new rules in order to make them applicable to either the first or the second constituent of the aggregated configuration. Generally,
{\small
\begin{equation*}
\begin{aligned}
\cell{cfg}{\cell{code}{c_1}\; \cell{env}{e_1}\; \cell{defs}{d_1}} \land \textit{cond} \Rightarrow \cell{cfg}{\cell{code}{c_2}\; \cell{env}{e_2}\; \cell{defs}{d_2}} \text{\normalsize\qquad turns into} \\
\bcell{eq}{\cell{cfg1}{\cell{code}{c_1}\; \cell{env}{e_1}\; \cell{defs}{d_1}}\; \ldots} \land \textit{cond} \Rightarrow \bcell{eq}{\cell{cfg1}{\cell{code}{c_2}\; \cell{env}{e_2}\; \cell{defs}{d_2}}\; \ldots}
\end{aligned}
\end{equation*}}
as well as a similar rule for $\cell{cfg2}{\ldots}$.

For expressing the partial equivalence, we have to define an RL formula of form $S_1 \Rightarrow S_2$, where $S_1$ represents the initial state of the two programs with possible conditions, whilst $S_2$ is a state pattern expressing the equivalence.

\paragraph{Symbolic Circular Coinduction.} A sound and relatively complete 
7-rule proof system is available~\cite{eq-corr} for RL, which is theoretically suitable for proving the correctness property expressing program equivalence. However, this 7-rule proof system is rather complex, there is no practical strategy published for building the proofs.

One of our goals is to have a semi-automatic system for verifying refactorings, so we have chosen a simplified version of the above mentioned proof system introduced in a related technical report~\cite{sym-ex}. \emph{Symbolic Circular Coinduction} (SCC) is coinduction-based extension of symbolic execution that can be used for deductive verification of program properties specified by RL formulas. The proof system consists of 3 inference rules, and can be easily implemented with a straightforward tactic. The report presents a prototype tool that can automatically build proofs if we give a language definition and an RL formula expressing some correctness property (see Section \ref{sec:auto}). Note that both the 7-rule and 3-rule proof systems are sound only on deterministic languages. For non-deterministic languages they introduced \emph{all-path reachability logic}~\cite{all-path-rl}, but it is future work to examine how to express equivalence in this logic.

\subsection{Local Refactorings}
\label{sec:ver-short}


In this section we show how to check the correctness of local refactoring definitions by constructing an RL formula expressing the equivalence of the matching and the replacement pattern under the given condition. We use the SCC proof system for verifying the RL formula.

We suppose that we have a set containing RL formulas that capture the semantics of Erlang. Let $\eqref{eq:st1} \Rightarrow \eqref{eq:st2}$ be the RL formula expressing the correctness property. We can mechanically construct the matching logic formula~\eqref{eq:st1} for a rule given in the form defined in Section \ref{sec:localdef}. We fill in the configuration specified for equivalence checking by putting matching and replacement patterns into the \textsf{code} cells of \textsf{cfg1} and \textsf{cfg2}, respectively. The metavariables of the patterns become mathematical variables of the formula.
As we would like to check whether the patterns are equivalent in any environment, we put new mathematical variables into both the \textsf{env} cells and the \textsf{defs} cells.
We append the condition to the configuration with logical conjunction.
\sms[eq:st1]{
 \bigcell{eq}{
 \begin{aligned}
 \cell{cfg1}
 {&\cell{code}{\texttt{<matching pattern>}}\;
 &&\cell{env}{e_1}\; &&\cell{defs}{d_1}\;} \\
 \cell{cfg2}
 {&\cell{code}{\texttt{<replacement pattern>}}\;
 &&\cell{env}{e_1}\; &&\cell{defs}{d_1}\;}
\end{aligned}
 }
 \land \texttt{<condition>} 
}%
The condition can contain various semantic functions and predicates. In order to be able to use them in the proof, we have to axiomatize them with RL formulas and add the axioms to the set of the formulas defining the object language. For example, the \texttt{fresh} predicate, specifying a variable name be fresh in the scope of the expression, is defined as follows:
\sms[eq:fresh]{
\bcell{eq}{\cell{cfg1}{\cell{env}{e_1}\; \ldots}\; \cell{cfg2}{\ldots}} \land \texttt{fresh}(x) \Rightarrow
\bcell{eq}{\cell{cfg1}{\cell{env}{e_1}\; \ldots}\; \cell{cfg2}{\ldots}} \land x \notin \texttt{keys}(e_1) \land \texttt{isVar}(x)
}%
Let us now define the formula expressing the equivalence relation.
We say that the two configurations (and therefore the original code patterns) are equivalent if we can reach a state in their symbolic evaluation, where the code cells have exactly the same content as well as the variable environments are equal. (We can ignore the function definitions as they cannot be changed with a rule applied to an expression.)
\sms[eq:st2]{
\bcell{eq}{
\cell{cfg1} {\cell{code}{c_2}\; \cell{env}{e_2}\; \ldots \;} \quad
\cell{cfg2} {\cell{code}{c_2}\; \cell{env}{e_2}\; \ldots \;}}
}%

\paragraph{Example proof.} We show a proof sketch for the \emph{extract\_listhead} (Listing~\ref{lst:extractlisthead}) refactoring definition. As a first step, we compose a formula expressing the equivalence of the matching and replacement code patterns of the rule: this is the initial goal of the proof.
\sms{
 \bigcell{eq}{
 \begin{aligned}
  \cell{cfg1}
   {&\codecell{[\;h\;|\;t\;]}\;
   &&\cell{env}{e_1}\; &&\cell{defs}{d_1}} \\
  \cell{cfg2}
   {&\codecell{\texttt{begin}\; v\; =\; h,\; [\;v\;|\;t\;]\ \texttt{end}}\;
   &&\cell{env}{e_1}\; &&\cell{defs}{d_1}} \\
 \end{aligned}
 }
 \land \texttt{fresh}(v) \Rightarrow \eqref{eq:st2}
}%
An inference rule of SCC allows us to apply any of the formulas of the language semantics as a rewrite rule on the left-hand side of our goal. After using \eqref{eq:r1}, performing a begin-end elimination and applying \eqref{eq:fresh} on the \textsf{cfg2} cell, we acquire the following new goal:
\sms{
 \bigcell{eq}{
 \begin{aligned}
  \cell{cfg1}
   {&\codecell{[\;h\;|\;t\;]}\;
   &&\cell{env}{e_1}\; &&\cell{defs}{d_1}\;} \\
  \cell{cfg2}
  {&\codecell{\texttt{case}\; h\; \texttt{of}\; v\; \texttt{->}\; [\;v\;|\;t\;]\; \texttt{end}}\;
  &&\cell{env}{e_1}\; &&\cell{defs}{d_1}\;}
 \end{aligned}
 }
 \land v \notin \texttt{keys}(e_1) \land \texttt{isVar}(v)
 \Rightarrow \eqref{eq:st2}
}%
Finally, after applying \eqref{eq:r2} on \textsf{cfg2} we get:
\sms{
 \bigcell{eq}{
 \begin{aligned}
 \cell{cfg1}
 {&\codecell{[\;h\;|\;t\;]}\;
 &&\cell{env}{e_1}\; &&\cell{defs}{d_1}\;} \\
 \cell{cfg2}
 {&\codecell{[\;h\;|\;t\;]}\;
 &&\cell{env}{e_1}\; &&\cell{defs}{d_1}\;}
 \end{aligned}
 } 
\Rightarrow 
 \underbrace{\bigcell{eq}{
 \begin{aligned}
 \cell{cfg1}
 {&\cell{code}{c_2}\;
 &&\cell{env}{e_2}\; &&\ldots \;} \\
 \cell{cfg2}
 {&\cell{code}{c_2}\;
 &&\cell{env}{e_2}\; &&\ldots \;}
 \end{aligned}
 }}_{\eqref{eq:st2}}
}%
In the acquired formula, the left-hand side implies the
right-hand side in the sense of matching logic, which is the axiom in the proof system, so our initial goal is proven.

\subsection{Refactoring Schemes}
\label{sec:ver-idiom}

In general, \emph{extensive} refactoring definitions 
cannot be automatically verified. However, in many cases, they can be split into two parts: 1) a mechanically verifiable generic transformation pattern, and 2) a specific instantiation that can be automatically verified. We call the former part the refactoring scheme, which is predefined, and it is proven to be correct with respect to an instantiation contract. On the other hand, the specific part is called the parametrisation, and it is automatically checked for conformance with the contract. In this section, we overview the contracts belonging to the schemes defined in Section~\ref{sec:schemedef} and we outline the technique we use to verify refactorings specified with schemes.

\subsubsection{Function signature refactoring}

Refactorings of this scheme change the signature of a function according to a simple rewrite rule. Both the definition as well as the referring expressions are to be adjusted, such that the data and control flow is not altered. Consequently, every mention of the refactored function has to point to the modified signature and has to maintain its effect. For instance, function calls have to expand to the same series of expressions as well as their parameters have to bind as before.


We assume that every syntactic element to be modified is identified by the static analysis framework (the proof of this is beyond the scope of the paper); thus, the verification of the scheme lies in showing that the various changes will be consistent. The name and the arguments are modified according to the same rule at all definition and reference sites, so we only have to make sure that the rule is generic enough to apply to all the change candidates, and it does not make the arguments any less or more general so that the data and control flow is preserved. This leads to the following contract:
\begin{enumerate}
\item The matching pattern only contains metavariables, and the pattern for the arguments is linear. This guarantees that the rule applies to all the definitions and references.
\item In the replacement pattern, the name is the same metavariable as in the matching pattern (holding the old name) or a constant (determining a new name). The arguments contain the metavariables present in the matching pattern, but no other metavariables or constants: this guarantees that the new signature is compatible with the old one in terms of matching generality (i.e. calls have the same formal to actual parameter assignments as earlier).
\end{enumerate}

\noindent
The contract is checked as follows. Suppose we have three operations on argument list patterns that preserve generality: a) swapping two elements, b) duplicating elements and c) grouping elements into lists or tuples. If the argument list's matching pattern can be transformed into its replacement pattern with the previous operations, they are compatible: applying the rule on the formal and actual parameters of a function results in the same matching (this can be proven with the definition of \verb|getMatching|).
 
\subsubsection{Dataflow refactoring}

Refactorings defined with the forward or backward dataflow schemes modify elements along dataflow paths. At least two rewrite rules have to be supplied as parameters: one for transforming data sources and one for changing references.

As in the previous scheme, we rely on the correctness of the dataflow analysis, and only verify whether the changes will be consistent if applied to all the data definitions and references on the flow. The contract in this case is simple: every combination of the definition and reference rules have to result in equivalent expressions.

The idea is based on the fact that the reference transformations are changing expressions where the value flows into, so that we might replace the value by its definition. Thus, for every pair of rules we can mechanically construct an equivalence problem by replacing every occurrence of the reference metavariable in the upper part of the reference rule by the upper part of the definition, and by replacing the occurrences of the same metavariable in the lower part of reference rule with the lower part of the definition. Observe that this equivalence problem can be proved exactly the same way as shown in Section \ref{sec:ver-short}.

\paragraph{Example.} From the rule of Listing \ref{lst:fwdata}, we acquire two equivalence formulas to be proven, which in this case can be easily done by relying on the semantics of function invocation:

{\small
\begin{equation*}
\begin{aligned}
\bcell{eq}{
\cell{cfg1} {\codecell{\texttt{fun() ->} \;x\; \texttt{end ()}}\; \cell{env}{e_1}\; \cell{defs}{d_1}\;} \;
\cell{cfg2} {\codecell{x}\; \cell{env}{e_1}\; \cell{defs}{d_1}\;}}
\land \texttt{pure}(x) \Rightarrow \eqref{eq:st2} \\
\bcell{eq}{
\cell{cfg1} {\codecell{\texttt{apply(fun() ->} \;x\; \texttt{end,[])}}\; \cell{env}{e_1}\; \cell{defs}{d_1}\;} \;
\cell{cfg2} {\codecell{x}\; \cell{env}{e_1}\; \cell{defs}{d_1}\;}}
\land \texttt{pure}(x) \Rightarrow \eqref{eq:st2}
\end{aligned}
\end{equation*}}%


\subsection{Concrete Application of Refactorings}
\label{sec:ver-app}

When we cannot verify the refactoring definition, we have the possibility to verify just one concrete application of the refactoring. We would like to check whether the original and the resulting code have the same behaviour, for one particular choice of refactoring and code. The difficulty with equivalence checking of Erlang programs is that we do not have any main function or expression. Instead, since all of the exported functions can be called from outside a module, we chose to check whether these functions have their behaviour preserved by the transformation.

We can mechanically prove this by collecting all of the exported functions of the modules both from the original and the transformed code, generate an RL formula for each pair of functions expressing their equivalence, and finally, prove all of the RL formulas with SCC or with other suitable proof system.

In the left-hand side of the RL formula there should be the \textsf{eq} configuration with function calls to the function under consideration with the same (symbolic) variables as arguments in both \textsf{code} cells, with all of the original and transformed function definitions in the \textsf{defs} cell, and with empty \textsf{env} cells. The right-hand side of the formula should express the equivalence criteria: codes have to be derived to the same (concrete or symbolic) value, and we do not care about \textsf{env} and \textsf{defs}, as there should not remain any variable or function call in the \textsf{code} cell.

\subsection{Semi-automatic Method}
\label{sec:auto}

The proposed proof system, SCC, has a prototype implementation~\cite{sym-ex}, that is an extension of the rewrite-based executable semantic framework called the $\mathbb{K}$ framework~\cite{k-web}. The parameters of a proof in the system are a $\mathbb{K}$ language definition and a RL formula given in a simple, XML-like syntax. $\mathbb{K}$ and matching logic fit well together, as RL formulas can be expressed as rewrite rules in  $\mathbb{K}$ and, additionally, the $\mathbb{K}$ framework offers many features that ease the definition of the semantics of a language. 

We have defined a sublanguage of Erlang in the $\mathbb{K}$ framework, and using the prototype version of SCC we have successfully specified and verified some of our simple refactorings automatically. Currently, we have to specify RL formulas for the refactoring definitions by hand, but as a future work, we plan to implement a translator for it. 


\section{Related Work}
\label{sec:related}

It is a widely applied technique to employ context-free conditional rewrite rules and functional strategies~\cite{Bravenboer200852} to implement program transformations. Bravenboer and Olmos show that by adding dynamically defined rewrite rules into the system~\cite{Bravenboer:2005:PTS:1227247.1227253}, context-dependent (even data flow driven) transformations~\cite{Olmos05} are definable; however, these definitions are hardly verifiable.

Effort has been put into formal specification, verification and implementation of refactorings. The fundamental work of Opdyke~\cite{Opdyke:1992:ROF:169783} suggests refactorings be composed of basic steps called microrefactorings. For object-oriented languages, Schaefer~\cite{Schaefer:2010:SIR:1932682.1869485} introduced a system in which he reasoned about semi-formal definitions of a set of basic refactorings, but the proofs are mostly informal. Roberts~\cite{Roberts:1999:PAR:929806} applies a different definition style, with an emphasis on the side-conditions and proper composition of the base refactorings. However, neither provides formally verified or executable definitions.

Semantics-aware, verifiable transformations can be specified by graph rewriting~\cite{SMR:SMR316} as well, but the resulting graphical descriptions are relatively complex compared to concrete syntax patterns. Padioleau et al.~\cite{Padioleau:plos06} propose a transformation language incorporating semantic conditions into the textual patterns; however, they use it for specifying patches rather than refactorings.
Verbaere~\cite{Verbaere:2006:JSL:1134285.1134311} proposes a compact, representation-level formalism for executable definitions; it is language-independent, but does not give support for verifying the correctness of refactorings. For Erlang, our previous work~\cite{tfp14} drafts a refactoring language, solely focusing on simplicity and interpretability. Also for Erlang, Li~\cite{Li:2012:DLS:2259278.2259323} defines an API for describing microrefactorings and a feature-rich language for composition, but formal verification is not addressed. There are some results~\cite{ver-ref-sultana} in defining provably correct refactorings for simple languages, but for real-world cases, the question is still open. Our work aims to take a significant step further, and offers not only refactoring-specific proofs, but a generic verification technique for custom-defined transformations.



\section{Discussion}

In Section~\ref{sec:defs} we have drawn the design goals of our refactoring definition approach: specifications should be \emph{executable}, \emph{verifiable}, \emph{intuitive} and \emph{applicable}. Let us discuss the limitations and future work related to each of these goals.

\emph{As for executability and verifiability.}
The methods and examples presented in the paper have been successfully implemented in our prototype. As \cite{Ekman:2008:RT:1636642.1636647} points out, refactoring consists of analysis plus transformation; we put the focus on transformation, whereas we made use of the analysis infrastructure present in RefactorErl. Consequently, even though we focussed on verifying the transformations, our transformations are only correct if the tree manipulation and static analysis framework (with the language meta-theory) is correct. In the future we anticipate broadening the scope of the work to include more constructs of Erlang, and to extend the verification capacity too.

\emph{As for applicability.}
Our formalism is intentionally restrictive: by providing a less general toolset to the refactoring programmer, we can give guarantees in return. For instance, the language lacks unbounded recursion, but this way termination is not an issue. We have demonstrated the applicability of our approach by defining well-known refactoring transformations, but it is to be investigated whether all the meaningful Erlang refactorings can be phrased in the language or we have to get rid of some constraints and give more control to the refactoring programmer. In order to make it more accessible, we plan to integrate our solutions into Erlang IDEs and provide a convenient interface for defining, checking and executing user-defined refactoring transformations.

\emph{As for intuitiveness.}
Formalising simple rewritings with semantic conditions is straightforward in the language, but specifying extensive refactorings with schemes is not always that obvious. Another nontrivial step in defining verifiable refactorings is decomposition: in case of complex transformations, it might take considerable effort to find smaller refactoring steps that composed together define the very same transformation. We will investigate tool support for inferring schemes in extensive steps as well as for giving hints regarding decomposition.

\section{Conclusions}
\label{sec:conclusion}

We have shown that it is possible to define a framework for describing refactorings for a particular programming language -- Erlang -- in such a way that the descriptions are high-level and readable, but at the same time they are executable. They are, moreover, amenable to verification using a rewriting logic framework, and in some cases verifications of refactorings, or of particular applications of them, are derivable automatically.

Our approach is, at some points, language-specific: the semantic predicates and functions are in line with the concepts of Erlang, and also, the high-level refactoring skeletons would probably be different in other languages. Nevertheless, we believe that the main idea would be adaptable in refactoring tools for other functional languages, and that the lesson of specialising the formalism to work smoothly with a single language will be equally valid for other languages, functional or otherwise.

\section{Acknowledgements}

We thank the anonymous reviewers for their valuable and constructive comments, which helped us  to improve this paper considerably.

We are grateful to Andrei Arusoaie and Dorel Lucanu for providing us with the pre-release copy of the SCC extension of $\mathbb{K}$ used to perform some of the verifications reported here.

This work has received funding from the European Institute of Innovation and Technology (EIT). This European body receives support from the Horizon 2020 research and innovation programme.

This work has been supported by the European Union Framework 7 under contract no. 288570. ParaPhrase: Parallel Patterns for Adaptive Heterogeneous Multicore Systems.

\label{sect:bib}
\bibliographystyle{eptcs}
\bibliography{vpt16}

\end{document}